\documentclass{ws-p8-50x6-00m}

\newcommand{\eg}{{\em e.g.}}
\newcommand{\ie}{{\em i.e.}}

\newcommand{\ccbar}{$c\bar{c}$}
\newcommand{\pythia}{{\sc Pythia}}
\newcommand{\atlas}{{\sc Atlas}}
\newcommand{\lhcb}{{\sc lhc}-b}
\newcommand{\pt}{$p_{\perp}$}
\newcommand{\etal}{{\it et al.}}
\newcommand{\jpsi}{$J/\psi$}
\newcommand{\psip}{${\psi}^{\prime} \,$}
\newcommand{\Y}{$\Upsilon \,$}
\def\EPJC{ {\em Eur. Phys. J.} C}
\newcommand{\lw}{\MakeLowercase}

\begin{document}

\title{Prompt $J/\psi$ production from Tevatron to LHC
 \thanks {P\lw{resented by} C. B\lw{renner} M\lw{ariotto} \lw{at the} VIII I\lw{nternational} W\lw{orkshop on} H\lw{adron} P\lw{hysics} (Hadrons 2002), B\lw{ento} G\lw{on\c{c}alves}, B\lw{razil}, 14 - 19 A\lw{pril} 2002}
}

\author{C.~Brenner~Mariotto$^{1,2}$, J.~Damet$^{2,3}$, G.~Ingelman$^{2,4}$}

\address{$^1$ Institute~of~Physics,~Univ.~Fed.~do~Rio~Grande~do~Sul,~Box~15051,~CEP~91501-960~Porto~Alegre,~Brazil\\ 
$^2$ High Energy Physics, Uppsala University, Box 535, S-75121 Uppsala, Sweden \\ 
$^3$ Laboratory for High Energy Physics, University of Bern, Bern,~Switzerland\\
$^4$ Deutsches~Elektronen-Synchrotron~DESY, 
Hamburg,~Germany\\
E-mail: mariotto@if.ufrgs.br}  

\maketitle

\abstracts{Models with essential non-perturbative QCD dynamics and describing Tevatron data on high-$p_\perp$ charmonium are extrapolated to give predictions of prompt $J/\psi$ production at the LHC. Differences of up to an order of magnitude occurs. An important point is here the treatment of higher order perturbative QCD effects. }

The interplay of hard and soft QCD effects has been demonstrated in Tevatron data \cite{Tevatron} for high-\pt \ \jpsi , \psip  and \Y , which are up to factors of 50 above the pQCD prediction in the Colour Singlet Model (CSM) \cite{csm}, where a colour singlet \ccbar \ pair is produced at the parton level. This deficit can be explained by letting a fraction of the more abundant colour octet \ccbar \ pairs be transformed into colour singlet through some soft QCD dynamics. This has been described in different models: the Colour Octet (COM) \cite{com}, Colour Evaporation (CEM) \cite{cem}, Soft Colour Interaction (SCI) \cite{sci} and Generalized Area Law (GAL) \cite{gal} models, which can be made to fit these Tevatron data. 

In this work we study the extrapolation of CEM, SCI and GAL models to the LHC energy and examine its theoretical uncertainty \cite{jpsilhc,san1}. These models are based on a similar phenomenological approach, where soft colour interactions can change the colour state of a \ccbar \ pair from octet to singlet. They employ the same hard processes to produce a \ccbar \ pair regardless of its spin state. Of importance for high-\pt \ $J/\psi$ production are NLO tree diagrams with a third hard parton that balances the \pt \ of the $c\bar{c}$ pair. The most important one is the process $gg\to c\bar{c}g$ with a gluon exchange in the $t$ channel, since the basic $2\to 2$ process $gg\to gg$ (followed by a gluon splitting $g\to c\bar{c}$) has a much larger ${\cal O}(\alpha_s^2)$ cross section than the LO $gg \rightarrow c\bar{c}$ and $q\bar{q} \rightarrow c\bar{c}$ production processes. Still higher orders than NLO can be important at the Tevatron and LHC energies, since many gluons can be emitted and their virtuality need not be very large to allow a split into a $c\bar{c}$ pair. Higher order processes can be approximately described by the parton shower approach available in the \pythia \ \cite{pythia} Monte Carlo, where in all basic QCD $2\to 2$ processes the incoming and outgoing partons may branch as described by the DGLAP equations. 

\begin{figure}[t]
\begin{center}
\epsfxsize=55mm 
\epsfbox{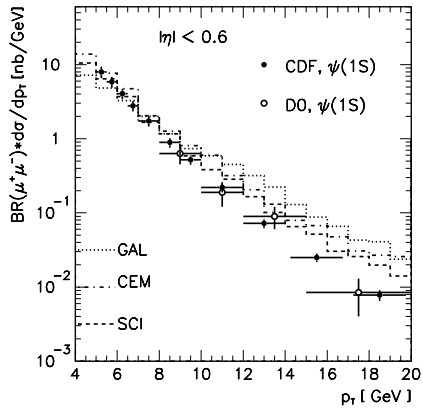}
\vspace{-0.15cm}
\caption{{\label{fig:galsci}}Distribution in transverse momentum of prompt $J/\psi$ as observed by CDF and D0 ${\scriptsize {}^1}$ in $p\bar{p}$ interactions at the Tevatron and obtained in the CEM, SCI and GAL models. }
\end{center}
\vspace{-0.55cm}
\end{figure}

In CEM \cite{cem,cemmar} the exchange of soft gluons is assumed to randomise the colour state, implying a probability $1/9$ that a $c\bar{c}$ pair is colour singlet and produces charmonium if its mass is below the threshold for open charm production, $m_{c\bar{c}}<2m_D$. The fraction of a specific charmonium state $i$, relative to all charmonia, is given by a non-perturbative parameter $\rho_{i}$ ($\rho_{J/\psi}=0.4-0.5$).

In SCI \cite{sci,gaps,sci-onium} it is assumed that colour-anticolour, corresponding to non-perturbative gluons, can be exchanged between partons emerging from a hard scattering and hadron remnants, leading to different topologies of the confining colour string-fields and thereby to different hadronic final states. The probability to exchange a soft gluon between parton pairs is given by a phenomenological parameter $R$. The mapping of $c\bar{c}$ pairs below the threshold for open charm production is based on spin statistics  resulting in a fraction of a specific quarkonium state $i$ with total angular momentum $J_i$ and main quantum number $n_i$ given by $f_i = \frac{\Gamma_i}{\sum_k \Gamma_k} $, where $\Gamma = (2J_i+1)/n_i$. 

In GAL \cite{gal}, a generalisation of the area law suppression $e^{-bA}$ (with $A$ the area swept out by the string in energy-momentum space) gives a dynamic probability $R=R_0(1-e^{-b\Delta A})$ for two string pieces to interact depending on the area difference $\Delta A$ resulting from the changed string topology. This favours shorter strings and thereby quarkonium production. The parameters $R_0$ and $b$ are obtained from a fit to both HERA and LEP data.

\begin{figure}[t]
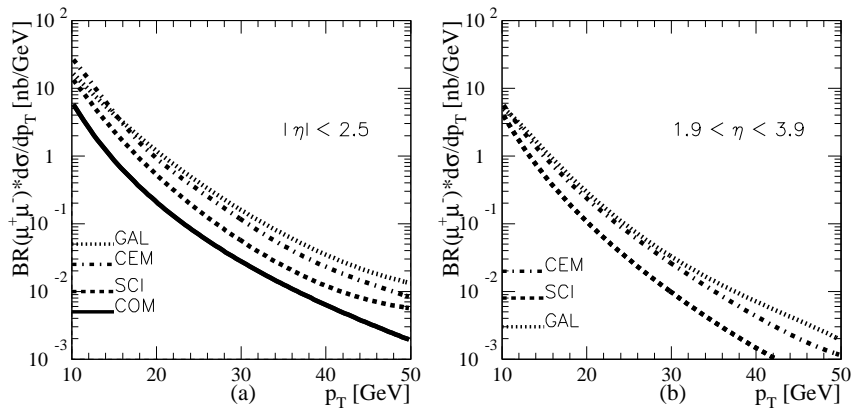

\parbox[t]{5.6cm} {
\epsfxsize=5.5cm
\epsffile{hadrons02-atlas.epsi}
}
\parbox[t]{5.6cm} {
\epsfxsize=5.5cm
\epsffile{hadrons02-lhcb.epsi}
}
\caption{\label{fig:lhc} Differential cross sections in transverse momentum for $J/\psi$ in $pp$ collisions at $\sqrt{s}=14$\ TeV based on the CEM, SCI and GAL models. In all cases the $J/\psi$ is required to have $p_\perp >10$\ GeV and decay into $\mu^+\mu^-$, required to be within the indicated rapidity coverage of the \atlas \ (a) and \lhcb \ (b) detectors. For comparison, the COM results from ${\tiny {}^8} $ are included in (a).}
\vspace{-0.2cm}
\end{figure}

The comparison of the CEM, SCI and GAL models with the Tevatron data is shown in Fig.~\ref{fig:galsci}. As can be seen, all models give a quite decent description of the data. Although the shape is not perfect in the tail of the distribution, it is quite acceptable given the simplicity of the models. The overall normalisation is correctly given by the models. For the CEM this is obtained by setting $\rho_{J/\psi}=0.43$ and the charm quark mass to 1.5 GeV. The SCI and GAL models have not been tuned to these data, but the result is also sensitive to the charm quark mass (taken as default $m_c=1.35$\ GeV in \pythia \ 5.7). The parton densities used are the default CTEQ3L \cite{CTEQ4L} for SCI and CTEQ4L \cite{CTEQ4L} for CEM and GAL. One should note that an arbitrary $K$ factor is not needed in any of the models, since higher order pQCD processes were included through the parton showers. Based on the ability of these models to correctly reproduce the Tevatron data and keeping all parameters fixed from this comparison, we can now extrapolate them to the LHC energy. 

Results of these models for LHC energies, \ie \ $\sqrt{s}=14$ \ TeV, are shown in Fig.~\ref{fig:lhc} for the acceptance regions of the future LHC experiments \atlas \ and \lhcb. For \atlas \ , the $p_\perp$ distribution in Fig.~\ref{fig:lhc}a is quite similar for the three models, although they differ somewhat in the high-$p_\perp$ tail. These predictions should, however, not be taken as very precise in view of the simplicity of these models that attempt to describe unknown non-perturbative QCD phenomena. The overall normalisation, which \eg \ is sensitive to the value of the charm quark mass, should not be considered to be better than within a factor two. For the \lhcb \ pseudorapidity cuts, SCI is suppressed relative to CEM and GAL, showing that the models have a somewhat different rapidity behaviour. 

We can also see in Fig.~\ref{fig:lhc}a that the COM result \cite{san1} is significantly lower than the other model results. As we have found recently \cite{jpsilhc}, part of this difference is because COM only includes $c\bar{c}$ pairs from the first $g\to c\bar{c}$ branching (namely a NLO approximation), whereas in our calculations with the CEM, SCI and GAL models, $c\bar{c}$ pairs are produced in any branching in the parton shower approximation, accounting for all higher orders. For the Tevatron the difference between these two approaches is within the precision of data, but for LHC the deviation can become as large as one order of magnitude. 

In conclusion, extrapolating models which can describe Tevatron data to the LHC energy, we find significant differences in the predicted prompt $J/\psi$ cross sections up to almost an order of magnitude. Part of this difference is related to the treatment of higher order contributions, which is important in order to obtain a good estimate of the correct prompt \jpsi \ production rate.

{\bf Acknowledgments:} 
This work was supported by the Swedish Research Council and 
by CAPES and CNPq, Brazil.

\end{document}